\documentclass[a4paper,12pt]{article}

\usepackage[T2A]{fontenc}
\usepackage[cp1251]{inputenc}
\usepackage[english]{babel}
\usepackage{graphicx}

\usepackage[centertags]{amsmath}
\usepackage{amsfonts}
\usepackage{amssymb}
\usepackage{amsmath}
\usepackage{newlfont}

\usepackage{multirow}
\usepackage{caption}
\usepackage{longtable}
\usepackage{array}
\usepackage{lscape}

\newlength{\defbaselineskip}
\newcommand{\setlinespacing}[1]%
           {\setlength{\baselineskip}{#1 \defbaselineskip}}

   \newtheorem{lemma}{Lemma}

\makeatletter
\def\redeflsection{\def\l@section{\@dottedtocline{1}{0em}{10em}}}
\renewcommand{\appendix}{\par%
  \setcounter{section}{0}%
  \setcounter{subsection}{0}%
  \renewcommand{\appendixname}{Attachment}%
  \def\sectionname{\appendixname}%
  \addtocontents{toc}{\protect\redeflsection}%
  \gdef\thesection{\@Alph\c@section\@arabic\c@table}%
}

\begin{document}

\title{\bfseries\boldmath Right sign of spin rotation operator}

\maketitle

\begin{center}
 \author{\bf R.\,A.~Shindin, D.\,K.~Guriev, A.\,N.~Livanov, I.\,P.~Yudin}
 \vskip 3mm
 {\small{\it VBLHEP JINR, 141980 Dubna, Russia}\\
 {\it E-mails: shindin@jinr.ru, gurdm@yahoo.com, livanov@jinr.ru, yudin@jinr.ru}}
\end{center}

\vskip 3mm

 \hfill Pacs: {25.40.Kv}

 \hfill UDC: {539.171.11}

\vskip 3mm


% ----------------------------------------------------------------

%\subjclass{}%

%\date{}%
%\dedicatory{}%
%\commby{}%
% ----------------------------------------------------------------

\if 1 
 \begin{abstract}
  В учебниках по квантовой механике
  при рассмотрении вопросов преобразования спина фермиона в пространстве
  предлагается использовать унитарный оператор 
  $\widehat{U}_{\vec n}(\varphi)=\exp{(-i\frac\varphi2(\widehat\sigma\cdot\vec n))}$,
  где $\varphi$ --- угол поворота вокруг оси $\vec{n}$.
  Однако этот оператор вращает спин в обратную сторону,
  т.е. представляет левовинтовое вращение.
  Ошибка интерпретации действия $\widehat{U}_{\vec n}(\varphi)$ возникает из-за того,
  что спин полагается обычным вектором,
  априори независимым от оператора $\widehat\sigma$.
  В работе показано, что каждый фермион, пусть его номер будет $i$,  
  имеет свой собственный Паули-вектор $\widehat\sigma_i$
  и преобразуется в пространстве вместе с ним.
  Если ввести глобальный оператор $\widehat\sigma$,
  т.е. определить одну общую для всех фермионов ось квантования $z$ 
  и работать в представлении сферы Блоха,
  то преобразования спина $\frac12$ формально не меняются:
  вращение вокруг $\vec{n}$ на угол $\varphi$
  по правилу правого винта
  задаётся сопряжённым оператором 
  $\widehat{U}^+_{\vec n}(\varphi)=\exp{(+i\frac\varphi2(\widehat\sigma\cdot\vec n))}$.
 \end{abstract}
\fi

\begin{abstract}
  For the fermion transformation in the space 
  all books of quantum mechanics propose to use the unitary operator  
  $\widehat{U}_{\vec n}(\varphi)=\exp{(-i\frac\varphi2(\widehat\sigma\cdot\vec n))}$,
  where $\varphi$ is angle of rotation around the axis $\vec{n}$.
  But this operator turns the spin in inverse direction
  presenting the rotation to the left.
  The error of defining of $\widehat{U}_{\vec n}(\varphi)$ action is caused 
  because the spin supposed as simple vector
  which is independent from $\widehat\sigma$-operator a priori.
  In this work it is shown that each fermion marked by number $i$
  has own Pauli-vector $\widehat\sigma_i$
  and both of them change together.
  If we suppose the global $\widehat\sigma$-operator and 
  using the Bloch Sphere approach
  define for all fermions the common quantization axis $z$
  the spin transformation will be the same:
  the right hand rotation around the $\vec{n}$-axis
  is performed by the operator 
  $\widehat{U}^+_{\vec n}(\varphi)=\exp{(+i\frac\varphi2(\widehat\sigma\cdot\vec n))}$.
 \end{abstract}

% ----------------------------------------------------------------

%\received{\formatdate{25}{3}{2015}}

%\maketitle

%\if 1 

\pagebreak
\section{Introduction}
   The spin-vector is an analog of mechanical moment
   but its projection to any direction has a discrete values.
   In the case of particle with spin $\frac12\hbar$
   the projection will be equal to $s_z=+\frac12\hbar$ or $-\frac12\hbar$.
   To account this duality 
   the three Hermitian $2\times2$ matrices are used 
   that forms the Pauli-operator $\widehat\sigma$:
  \begin{eqnarray}
   & \widehat\sigma = \vec{i}\sigma_x + \vec{i}\sigma_y
   + \vec{k}\sigma_z \;,\quad\textrm{where} 
     \label{theory.Spin Pauli-operator} \\ %[2mm]
   & \sigma_x = \begin{pmatrix}0&1\\1&0\end{pmatrix}\;,\quad
     \sigma_y = \begin{pmatrix}0&-i\\i&0\end{pmatrix}\;,\quad
     \sigma_z = \begin{pmatrix}1&0\\0&-1\end{pmatrix}\;.
     \label{theory.Spin Pauli matrices}
  \end{eqnarray}

  Since only $\sigma_z$ has a diagonal view 
  its own vectors or spinors are pure states:  
  \begin{eqnarray}\label{theory.pure spinors}
   & \chi_z(s_z=+\frac12\hbar)=\dbinom10\;,\quad
     \chi_z(s_z=-\frac12\hbar)=\dbinom01\;.
  \end{eqnarray}

\section{Eigenvectors of operator $(\widehat\sigma\cdot\vec{r}\,)$}\label{Appendix.own_values_of_Pauli_operators}
\begin{figure}[!ht]
   \quad\centering
   \scalebox{.3}{\includegraphics{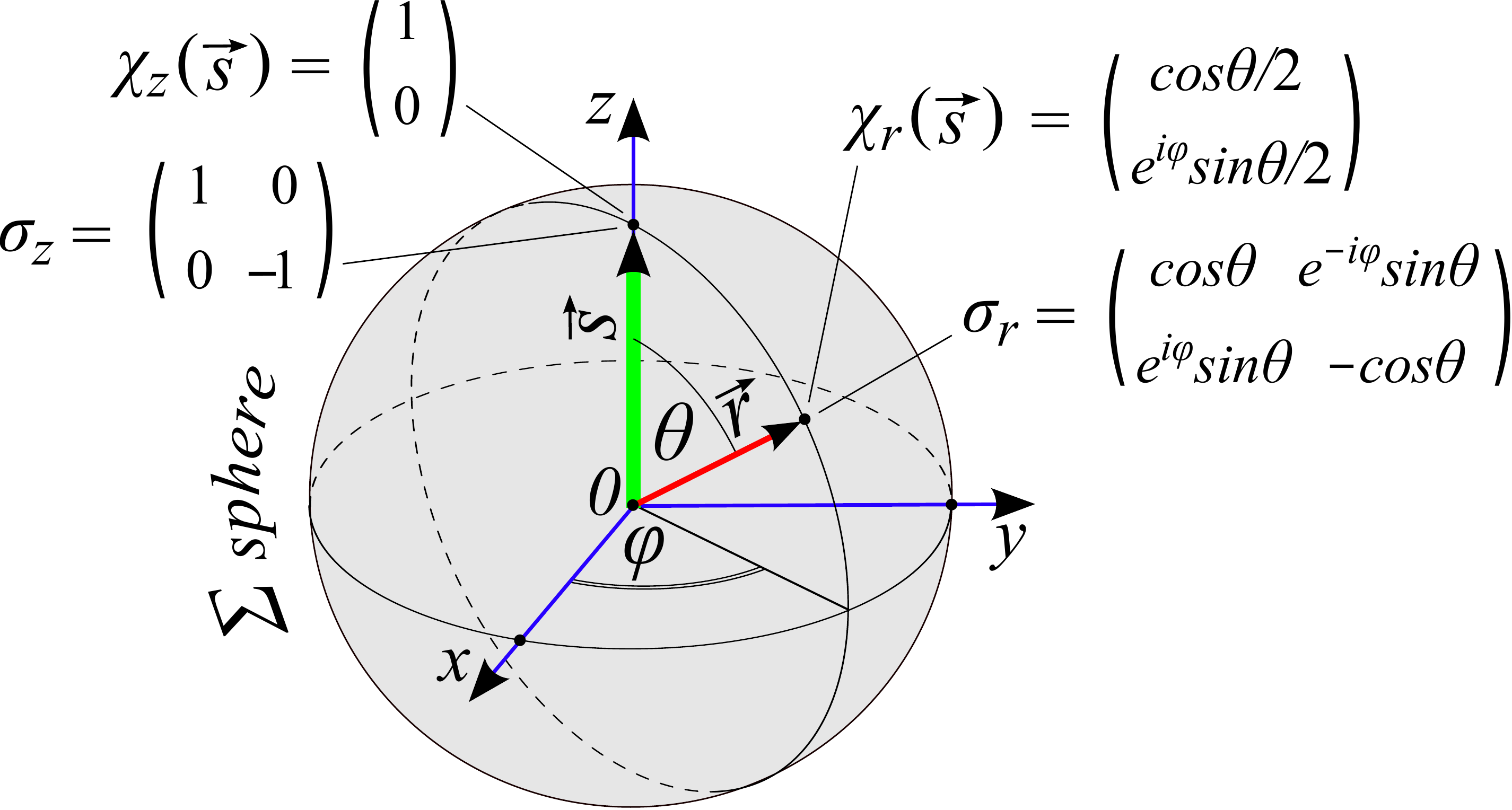}}\quad
   \caption{\small{Absolute value of projection of the Pauli-vector 
    $\widehat\sigma=\vec{i}\sigma_x+\vec{j}\sigma_y+\vec{k}\sigma_z$
    to any direction $\vec{r}$ equals to unit: $|\sigma_r|^2=1$.
    That allows to present it like a $\Sigma$-sphere.
    Each point of its %the $\Sigma$-sphere corresponds to the operator $\sigma_r$ which 
    has two eigenvectors or spinors $\chi_r(s_z=+\frac12\hbar)$
    and $\chi_r(s_z=-\frac12\hbar)$.
   }}\label{theory.Pauli-operator sphere}
  \end{figure}

 Let the unit vector $\vec{r}$ is defined in the spherical coordinate system
 by the values of zenith and azimutal angles $\theta$ and $\varphi$
 (Fig.~\ref{theory.Pauli-operator sphere}).
  
 In this case the scalar multiplying of $\widehat\sigma$ and $\vec{r}$ has the next form:
 \begin{eqnarray}
   &\sigma_r \;=\; (\widehat\sigma\cdot\vec{r}\,) \;=\;
   \sigma_x\sin\theta\cos\varphi+\sigma_y\sin\theta\sin\varphi+\sigma_z\cos\theta\;=\quad\nonumber\\ [2mm]
   &=\; \left(
         \begin{array}{cc}
           \cos\theta &  e^{-i\varphi}\sin\theta\\ [2mm]
           e^{i\varphi}\sin\theta & -\cos\theta \\
         \end{array}
       \right)
   \;=\; \left(
         \begin{array}{cc}
           \cos^2\frac\theta2-\sin^2\frac\theta2 &  2e^{-i\varphi}\sin\frac\theta2\cos\frac\theta2\\ [2mm]
           2e^{i\varphi}\sin\frac\theta2\cos\frac\theta2 & -\cos^2\frac\theta2+\sin^2\frac\theta2 \\
         \end{array}
       \right) \;=\quad\nonumber\\ [2mm]
   &=\; \chi_{(+)}\otimes\chi_{(+)}^+
   \;-\;\chi_{(-)}\otimes\chi_{(-)}^+ \;,
   \label{appendix.tensor view of Pauli-operator}\\ [2mm]
   &\chi_{(+)}\,=\,\dbinom{\cos\frac\theta2}{e^{i\varphi}\sin\frac\theta2}\;, \quad
    \chi_{(-)}\,=\,\dbinom{-\sin\frac\theta2}{e^{i\varphi}\cos\frac\theta2}\;.
   \label{appendix.own_values_of_Pauli_operators_anlong_any_direction}
 \end{eqnarray}
 Vectors $\chi_{(+)}$ and $\chi_{(-)}$ are unit and orthogonal among themselves:
 \begin{equation}
   \left|\chi_{(\pm)}\right|^2 = 1\quad,\qquad
   \chi^+_{(-)}\cdot\chi^{\phantom +}_{(+)} = 0\;.
 \end{equation}
 According to \eqref{appendix.tensor view of Pauli-operator} it is obviously:
 \begin{equation}
  \sigma_r\chi_{(+)} = \chi_{(+)}\quad,\qquad
  \sigma_r\chi_{(-)} = -\chi_{(-)}\,.
 \end{equation}

 Common phase of spinor's elements is not informative 
 therefore $\chi_{(-)}\equiv-\chi_{(-)}$.
 If the spin projection to the axis $z$ is positive $s_z=+\frac12\hbar$
 it can be presented as superposition of two states
 $s_r=+\frac12\hbar$ and $-\frac12\hbar$
 and their amplitudes equal to $\cos\frac\theta2$ and $e^{i\varphi}\sin\frac\theta2$ respectively.
 If the spin projection is negative $s_z=-\frac12\hbar$
 the states $s_r=+\frac12\hbar$ and $-\frac12\hbar$
 have amplitudes $-\sin\frac\theta2$ and $e^{i\varphi}\cos\frac\theta2$.
 
 The Pauli-operator along the vector $\vec{r}$
 can be associated with the Stern-Gerlach device\footnote{The difference
 between them is concluded in fact 
 that the Stern-Gerlach device measures not amplitudes 
 but the probabilities of states $\uparrow$ and $\downarrow$
 therefore the information about their relative phase
 $e^{i\varphi}$ is lost.}
 as it is considered in the Feynman Lectures \cite{Feynman-eng} (V.\,III, Ch.\,5,~6).
 The main of this device --- quantization axis. 
 When it coincides with the vector $\vec{r}$
 the device rotation around $\vec{r}$
 does not change the probability of spin states $s_r=+\frac12\hbar$ and $-\frac12\hbar$.
 The action of operator $\sigma_r$ is analogical ---
 from all possible representations of spin\footnote{In the SU(2) algebra 
 the definition of {\it pure} states is conditional
 because any representation of spin gives the full knowledge
 about its orientation in the space.
 If the $\chi_{(+)}$ and $\chi_{(-)}$ 
 \eqref{appendix.own_values_of_Pauli_operators_anlong_any_direction}
 will be named as {\it true}
 then any others can be defined using their linear combinations.
 For example:
 \begin{equation}
  \cos{\frac\theta2}\,\chi_{(+)}-\sin{\frac\theta2}\,\chi_{(-)}=\binom10\;,\quad
  \sin{\frac\theta2}\,\chi_{(+)}+\cos{\frac\theta2}\,\chi_{(-)}=e^{i\varphi}\binom01\equiv\binom01\;.
 \end{equation}}
 it saves unchanged only own spinors $\chi_{(+)}$ и $\chi_{(-)}$
 \eqref{appendix.own_values_of_Pauli_operators_anlong_any_direction},
 i.e.~only spin states along the vector $\vec{r}$.

\pagebreak 
\section{Rotation of spin $\frac12$ around an arbitrary axis}\label{Appendix.rotation of Pauli-operator}
\begin{figure}[!ht]
     \centering
     \scalebox{.4}{\includegraphics{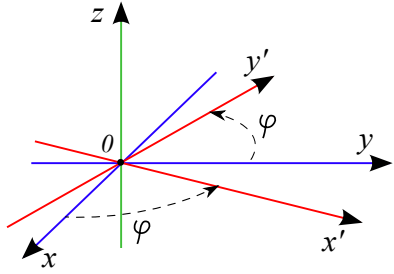}}\quad
     \caption{\small{The rotation of coordinate system around the axis $z$.
     }}\label{around z}
 \end{figure}

  Consider a coordinate system rotation $(x,\,y,\,z)\to(x',\,y',\,z)$
  around the axis $z$ by an angle $\varphi$ according to the right hand screw rule (Fig.~\ref{around z}).

 The matrix of transition has view:
   \begin{equation}\label{space around z}
      A=\left(\begin{array}{ccc}
        \cos\varphi & \sin\varphi & 0 \\
       -\sin\varphi & \cos\varphi & 0 \\
        0 & 0 & 1
      \end{array}\right):\quad
      A\left(\begin{array}{c}
        \vec{i} \\
        \vec{j} \\
        \vec{k}
      \end{array}\right)=
      \left(\begin{array}{c}
        \vec{i}\,' \\
        \vec{j}\,' \\
        \vec{k}
      \end{array}\right).
   \end{equation}

  Let the condition $\widehat\sigma=\textrm{const}$,
  i.e.~the spin Pauli-operator remains unchanging.
  In the system $(x',\,y',\,z)$ it can be presented as following:
   \begin{eqnarray}
        \widehat\sigma &=& \vec i\,'(\sigma_x\cos\varphi+\sigma_y\sin\varphi)+
        \vec j\,'(-\sigma_x\sin\varphi+\sigma_y\cos\varphi)+\vec k\sigma_z\;.
   \end{eqnarray}

  On the other hand the new coordinates $\sigma_{x'}$ and $\sigma_{y'}$
  can be defined by the unitary $2\times2$ matrices $u$ and $u^+$:
   \begin{eqnarray}\label{appendix.unitary matrix of Pauli-operator rotation}
    & \sigma_{x'} = u\,\sigma_x\,u^+ =
    \left(\begin{array}{cc}
        0 & e^{-i\varphi} \\
        e^{i\varphi} & 0 \\
    \end{array}\right),\quad
    \sigma_{y'} = u\,\sigma_y\,u^+ =
    \left(\begin{array}{cc}
        0 & -ie^{-i\varphi} \\
        ie^{i\varphi} & 0 \\
    \end{array}\right).
  \end{eqnarray}
  The solutions of equations 
  \eqref{appendix.unitary matrix of Pauli-operator rotation} are:
    \begin{equation}\label{appendix.su2 around z}
      u\equiv u_z=\pm\left(\begin{array}{cc}
         e^{-i\tfrac{\varphi}{2}} & 0 \\
         0 & e^{i\tfrac{\varphi}{2}}
      \end{array}\right),\quad
      u^+\equiv u^+_z=\pm\left(\begin{array}{cc}
         e^{i\tfrac{\varphi}{2}} & 0 \\
         0 & e^{-i\tfrac{\varphi}{2}}
      \end{array}\right).
   \end{equation}
  The marker $z$ says here 
  about the identical transformation $\sigma_z = u\,\sigma_z\,u^+$
  because this rotation carries out around the axis $z$.
  Without damage of our solution
  choosing in \eqref{appendix.su2 around z} the sign ``+''
  the unitary matrix $u_z$ can be rewritten as:
    \begin{equation}\label{appendix.su2 around z represented}
        u_z=\left(\begin{array}{cc}
                    1 & 0 \\
                    0 & 1
            \end{array}\right)\cos\frac{\varphi}{2}
          -i\left(\begin{array}{cc}
                    1 & 0 \\
                    0 & -1
            \end{array}\right)\sin\frac{\varphi}{2}=
        E\cos\frac{\varphi}{2}-i\sigma_z\sin\frac{\varphi}{2}\;.
    \end{equation}

  The formula \eqref{appendix.su2 around z represented} propogates
  to the case of system rotation around an arbitrary axis $\vec{n}$
  that leads to the unitary operator:
     \begin{equation}\label{appendix.opetrator of spin rotation}
        \widehat{U}_{\vec n}(\varphi)=
        E\cos\frac{\varphi}{2}-i(\widehat\sigma\cdot\vec n)\sin\frac{\varphi}{2}
        =\exp\Bigl[-i\frac\varphi2(\widehat\sigma\cdot\vec n)\Bigr]\;,
     \end{equation}
     \begin{equation}\label{appendix.transition of Pauli-operator}
        \widehat{U}_{\vec n}(\varphi)\;(\sigma_x,\,\sigma_y,\,\sigma_z)\;
        \widehat{U}^+_{\vec n}(\varphi)\;=\;(\sigma_{x'},\,\sigma_{y'},\,\sigma_{z'})\;.
     \end{equation}

 The transformation \eqref{appendix.transition of Pauli-operator}
 allows to express the coordinates of vector $\widehat\sigma$
 in the system $(x',\,y',\,z')$
 obtained from initial system $(x,\,y,\,z)$
 by the rotation around the vector $\vec{n}$ by an angle $\varphi$
 according to the right hand screw rule.
 For observer related with new system $(x',\,y',\,z')$
 it will be look like the left hand rotation around $\vec{n}$ by an angle $-\varphi$.
 Since the fermion states with pure projections $+\frac12\hbar$ and $-\frac12\hbar$
 are connected with old axis $z$ by the matrix $\sigma_z$ (hold by its dioganal form)
 this rule will be true for the spin too:

  \begin{lemma}[Spin rotation]\label{appendix.spin rotation around n.lemma}
  \setlinespacing{1.66}
  The rotation of the spin $\frac12$ around an arbitrary axis $\vec{n}$
  by an angle $\varphi$ corresponds to the operator
  $\widehat{U}^+_{\vec n}(\varphi)=\exp\Bigl[+i\dfrac\varphi2(\widehat\sigma\cdot\vec n)\Bigr]$.
 \end{lemma}

 Let us to consider the rotation around the axis $z$ by the angle $90^{\,\circ}$:
 \begin{eqnarray}\label{appendix.matrix rotation by 90 degrees around z}
  & \widehat{U}^+_z \equiv \widehat{U}^+_z(\frac\pi2) =
    \dfrac1{\sqrt{2}}\left(E+i\sigma_z\right) =
    \dfrac{1+i}{\sqrt{2}}
    \begin{pmatrix}
      1 & 0 \\
      0 & -i
    \end{pmatrix}, \\
  & \sigma_y \to \widehat{U}^+_z\,\sigma_y\,\widehat{U}_z = 
    \begin{pmatrix}
      1 & 0 \\
      0 & -i
    \end{pmatrix}
    \begin{pmatrix}
      0 & -i \\
      i & 0
    \end{pmatrix}
    \begin{pmatrix}
      1 & 0 \\
      0 & i
    \end{pmatrix} =
    \begin{pmatrix}
      0 & 1 \\
      1 & 0
    \end{pmatrix} = \sigma_x\;.
 \end{eqnarray}
 The projection $\sigma_y$ %along the vector $\vec{j}$ 
 is replaced by the matrix $\sigma_x$.
 It means that the Pauli-vector turns around the axis $z$ to the right.
 Since the spin state $\chi_y = \frac1{\sqrt{2}}\binom1i$ 
 is eigenvector of matrix $\sigma_y$
 the same changing is performed with the fermion:
 \begin{eqnarray}\label{appendix.spin by 90 degrees around z}
    \chi_y \to \widehat{U}^+_z\,\chi_y = 
    \frac{1+i}{\sqrt{2}}
    \begin{pmatrix}
      1 & 0 \\
      0 & -i
    \end{pmatrix}
    \frac1{\sqrt{2}}\binom1i = 
    e^{i\tfrac\pi4}\,\frac1{\sqrt{2}}\binom11 \equiv \frac1{\sqrt{2}}\binom11 = \chi_x\;.
 \end{eqnarray}
 If this transformation $\widehat{U}^+_z\,\chi_y=\chi_x$
 would be defined as the left rotation
 the connection between $\widehat\sigma$-operator and our spin would be broken
 converting it to the element of external space
 such as ort-vectors $\vec{i}$, $\vec{j}$, $\vec{k}$
 as a minimum or we need to suppose 
 that the spin and Pauli-operator are rotated in opposite directions. 
 
\pagebreak
 \begin{lemma}[Common moving]\label{appendix.common moving of spinors and Pauli-vector.lemma}
  \setlinespacing{1.66}
  The $\chi_r$ and $\sigma_r$ are changed identical.
 \end{lemma}
  Let us to take the $\sigma_r$ in the form 
  \eqref{appendix.tensor view of Pauli-operator}. Then:   
  \begin{eqnarray}
     &  u\sigma_ru^+ =\, u(\widehat\sigma\cdot\vec{r}\,)u^+ =\,
        u\chi_{(+)}\otimes\chi_{(+)}^+u^+ -\,
        u\chi_{(-)}\otimes\chi_{(-)}^+u^+ =  \nonumber \\ [2mm]
     &  =\, \left(u\chi_{(+)}\right)\otimes\left(u\chi_{(+)}\right)^+ -\,
        \left(u\chi_{(-)}\right)\otimes\left(u\chi_{(-)}\right)^+ =\,
        (\widehat\sigma'\cdot\vec{r}\,) \,=\, \sigma_r'\;.
        \label{appendix.common moving of spinors and Pauli-vector}
  \end{eqnarray}
  This transition $\sigma_r\to\sigma_r'$ is the transformation 
  of eigenvectors $\chi_{(\pm)}$ of operator $\sigma_r$
  to eigenvectors $\chi_{(\pm)}'=u\chi_{(\pm)}$ of operator $\sigma_r'=u\sigma_ru^+$,
  
  \hfill quod erat demonstrandum.

\bigskip
  The spin is related with the $\widehat\sigma$-operator
  and both of them are changed in the space together.
  In particular the state with pure projection $s_z=+\frac12\hbar$ 
  is rotated to the same direction like the $\sigma_z$ matrix.
  
\bigskip
  The transformation $\widehat\sigma\to\widehat\sigma'$ can be defined also
  by the $3\times3$ matrix $\widehat\sigma'=A\widehat\sigma$
  and for scalar multiplying $(\widehat\sigma'\cdot\vec{r}\,)$ 
  it gives the following:
  \begin{equation}
   (\widehat\sigma'\cdot\vec{r}\,) = (A\widehat\sigma\cdot\vec{r}\,) =
   (\widehat\sigma\cdot A^T\vec{r}\,) = (\widehat\sigma\cdot \vec{r}\,')\;.
  \end{equation}
  If the $A$-matrix in the system $(x,\,y,\,z)$ 
  rotates the vector $\widehat\sigma$ to the right
  it will be equal to the left rotation of vector $\vec{r}\to\vec{r}\,'$
  in the own coordinates of Pauli-operator.
  As it was shown before, 
  the direction of vector $\vec{r}$ on the $\Sigma$-sphere
  (Fig.~\ref{theory.Pauli-operator sphere}) 
  corresponds to the orientation of Stern-Gerlach device
  relative the fermion.

\pagebreak
\section {Common approaches to the determination \\ of spin rotation operator} 
 In all books of quantum mechanics
 the space rotation of spin has inverse definition of operator sign.
 For example in 8-th volume of Feynman's Lectures \cite{Feynman-eng} (V.\,III, Ch.\,6-3, P.\,6-15)
 we can find the analitical approach:
 if the $xy$-plane rotates around the axis $z$ by the angle $180^{\,\circ}$
 the transformation of pure states $s_z=+\frac12\hbar$ and $-\frac12\hbar$ %$\uparrow$ и $\downarrow$
 should add to them the phases $e^{im\pi}$ and $e^{-im\pi}$ respectively.
 Feynman chooses the $m=+\frac12$
 and in the same time he talks about second solution $m=-\frac12$
 that gives our definition of matrix $u_z$ \eqref{appendix.su2 around z}.
 The main Feynman's goal --- to show the difference of two phases 
 but it does not matter whether ``$+$'' or ``$-$''
 since the relative shift between them will be equal to $e^{i\pi}=-1$ anyway.

 \bigskip
 The standard solution can be find
 in the book of ``Theoretical physics'' \cite{Levich-eng}
 and here we consider the main points of this method.
 Supposing that new system $(x',\,y',\,z')$ is produced
 from $(x,\,y,\,z)$ using rotation around the axis $z$ by small angle $\delta\varphi$
 (Fig.~\ref{around z}, Form.~\ref{space around z}).
 If the quantum object $\psi(x,\,y,\,z)$ states unchanging
 then its representation in new system should be the same:
 \begin{eqnarray}\label{appendix.rule of any rotation}
   &  \cos\delta\varphi\approx1\;,\quad
      \sin\delta\varphi\approx\delta\varphi\;,\nonumber \\ [2mm]
   &  \psi(x,\,y,\,z) \equiv \psi'(x',\,y',\,z') = \psi'(x+y\,\delta\varphi,\,-x\,\delta\varphi+y,\,z)\;= \nonumber \\ [2mm]
   &  =\;\psi'(x,\,y,\,z)
      + y\,\delta\varphi\,\dfrac{\partial\psi'}{\partial x} - x\,\delta\varphi\,\dfrac{\partial\psi'}{\partial y}\;= \nonumber \\
   &  =\;\left[1-\delta\varphi\left(x\,\dfrac{\partial}{\partial y}-y\,\dfrac{\partial}{\partial x}\right)\right]
      \psi'(x,\,y,\,z)\;= \nonumber \\
   &   =\;\Bigl[1-\dfrac{i}{\hbar}\,\delta\varphi\,\hat{l}_z\Bigr]\psi'(x,\,y,\,z)\;=
      \;\exp{\Bigl[-\dfrac{i}{\hbar}\,\delta\varphi\,\hat{l}_z\Bigr]}\psi'(x,\,y,\,z)\;.
 \end{eqnarray}
 This rule \eqref{appendix.rule of any rotation}
 is propogated to the case of rotation
 around an arbitrary axis $\vec{n}$.
 Sum of rotations $\sum\delta\varphi_i=\varphi$ gives the formula:
 \begin{equation}\label{appendix.opetrator of any object rotation}
    \psi'(x,\,y,\,z)=\exp\Bigl[+\frac{i}{\hbar}\,\varphi\,\hat{l}_n\Bigr]\psi(x,\,y,\,z)\;.
 \end{equation}
 The projection of angular momentum operator to the direction $\vec{n}$
 is replased by the Pauli-operator
 and the wave function replaced by spinor:
 \begin{equation}\label{appendix.opetrator of spin rotation by book}
   \left.
    \begin{array}{c}
      \hat{l}_n\to\frac12\hbar\sigma_n \\
      \psi\to\chi
    \end{array}
   \right\}\quad\Rightarrow\quad   
    \chi'=\exp\Bigl[+i\frac\varphi2(\widehat\sigma\cdot\vec n)\Bigr]\chi\;.
 \end{equation}
 Since the coordinate system is rotated around $\vec{n}$ by the angle $\varphi$
 the observer related with new system $(x',\,y',\,z')$
 will see it as a rotation to the angle $-\varphi$,
 i.e.~the formula \eqref{appendix.opetrator of spin rotation by book}
 should produce the rotation of spin-vector to the left
 that contradicts with lemma \ref{appendix.spin rotation around n.lemma}.
 Performed replacments in \eqref{appendix.opetrator of spin rotation by book}
 can be justified in the sense that both spinor $\chi$ and operator $\sigma_n$
 belong to the SU(2) algebra (form it).
 However the $\chi$ and $\sigma_n$ are entered to this formula independent among themselves,
 i.e.~the spin behavior is not defined from the spin nature
 but are a kind of agreement.
 The problem is occurred also
 when in \eqref{appendix.rule of any rotation}
 we take the quantum mechanic definition 
 of angular momentum operator $\hat{l}_z$.
 Because in the Schr\"{o}dinger and Klein-Gordon equations
 the quadratic form $\nabla^2$ is used
 the sign of particle momentum is free:
 \begin{equation}
  \frac{\partial}{\partial x}\,\psi = \pm \frac{i}{\hbar}\,p_x\,\psi\quad
  \textrm{etc.}
 \end{equation}
 Choosing between ``$+$'' and ``$-$'' is a convention
 what wave function from $\exp\left[{\frac{i}{\hbar}\vec{p}\vec{r}\,}\right]$
 or $\exp\left[{-\frac{i}{\hbar}\vec{p}\vec{r}\,}\right]$ 
 has been taken.
 But in \cite{Levich-eng} this question is considered in a general view
 therefore the sign of operator $(\sigma\cdot\vec{n})$ %$\hat{l}_z$ 
 in the \eqref{appendix.opetrator of spin rotation by book}
 can be any
 and the direction of rotation around the vector $\vec{n}$
 is not defined at all.

 \bigskip
 In the <<Course of Theoretical Physics>>
 \cite{Landafshic-eng} (V.\,III, $\S$\,58) the rotation around the axis $z$
 is discribed by the matrix $\widehat{U}_z(\alpha)$ which coincides
 with the operator of lemma~\ref{appendix.spin rotation around n.lemma}.
 However in the text it does not define exactly 
 whether it is rotation of coordinate system or rotation of spin.
 Second rotation is performed around the axis {\it ON} 
 (around {\it knots axis} according to Euler)
 but defined using the operator $\widehat{U}_y(\beta)$
 that leads to an ambiguous definition for old and new coordinate systems.
 Therefore the $\widehat{U}_z(\alpha)$ 
 should be understood in the sense of rotation of spin vector.
 The next transformation $\widehat{U}_z(\gamma)$
 is supposed as a rotation around new axis $z$.
 This sequence of three rotations 
 $\widehat{U}_z(\gamma)\widehat{U}_y(\beta)\widehat{U}_z(\alpha)$
 which is shown on Fig.~20 in \cite{Landafshic-eng}
 can have only one interpretation --- 
 the spin is rotated together with its Pauli-vector
 as if repelled from the external space.
 In inverse case if the operator Pauli is fixed 
 in the initial coordinate system $(x,\,y,\,z)$
 these rotations should be performed around the axes $z$, $y'$ и $z''$,
 i.e.~total operator must be written as
 $\widehat{U}_{z''}(\gamma)\widehat{U}_{y'}(\beta)\widehat{U}_z(\alpha)$
 regardless from the question what kind of rotation 
 (right or left) is given by each matrix.
 If the transition into new system need to be done
 by the rotations around the axes of old basic
 then the matrices order should be inverse
 $\widehat{U}_z(\alpha)\widehat{U}_y(\beta)\widehat{U}_z(\gamma)$.

\pagebreak 
\section{Own space of spin}\label{Ch. Own space}
 Wrong interpretation 
 of operator \eqref{appendix.opetrator of spin rotation}
% to the fermion state 
 is related with initial supposing
 that this operator acts to the spin as external
 but it is not true.
 According to \eqref{appendix.tensor view of Pauli-operator}
 all three Pauli-matricies can be presented like a tenzorial multiplying
 of orthogonal spin states among themselves:
 \begin{subequations}\label{new definition of Pauli matrices}
  \begin{eqnarray}
   \sigma_z &=& \uparrow\otimes\uparrow^+ - \;\downarrow\otimes\downarrow^+,  \\ [2mm]
   \sigma_x &=& \uparrow\otimes\downarrow^+ + \;\downarrow\otimes\uparrow^+,  \\ [2mm]
   i\sigma_y &=& \uparrow\otimes\downarrow^+ - \;\downarrow\otimes\uparrow^+, \\ [2mm]
    E &=& \uparrow\otimes\uparrow^+ + \;\downarrow\otimes\downarrow^+.
  \end{eqnarray}
 \end{subequations}
 Unit matrix $E$ is added here for complete set. %of all possible combinations.
 By these definitions the matricies $\sigma_x$,  $\sigma_y$ and $\sigma_z$
 become Hermitian and unitary automaticaly
 and satisfy to the commutation relations:
  \begin{eqnarray*}
   \sigma_z = \sigma_z^+\,,\quad
   \sigma_z^2 = E\,,\quad
   \sigma_z\sigma_x = - \sigma_x\sigma_z = i\sigma_y\,\quad
   \textrm{etc.}
  \end{eqnarray*}
 
\begin{figure}[!ht]
     \centering
      \scalebox{.2}{\includegraphics{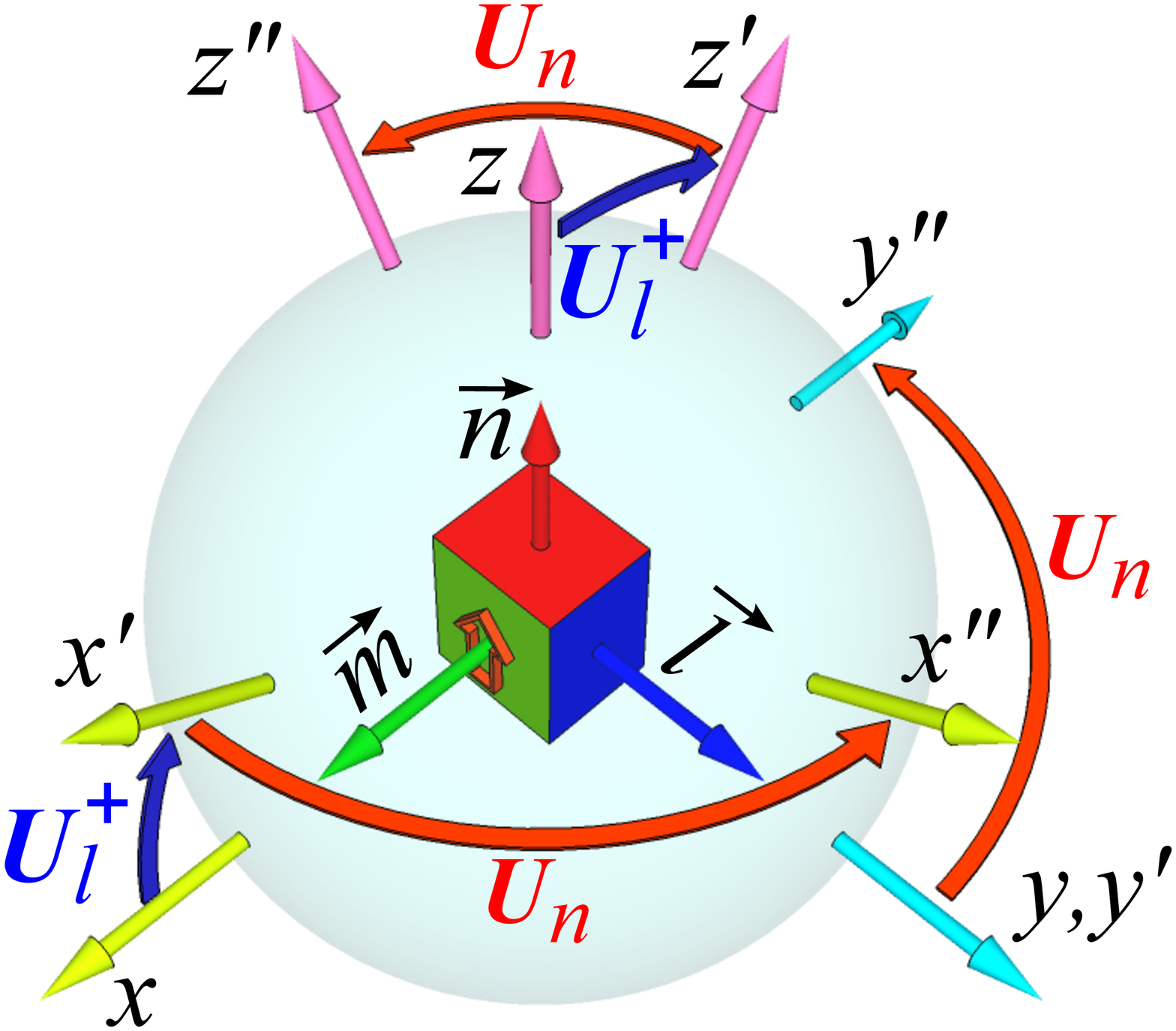}}\qquad
      \scalebox{.2}{\includegraphics{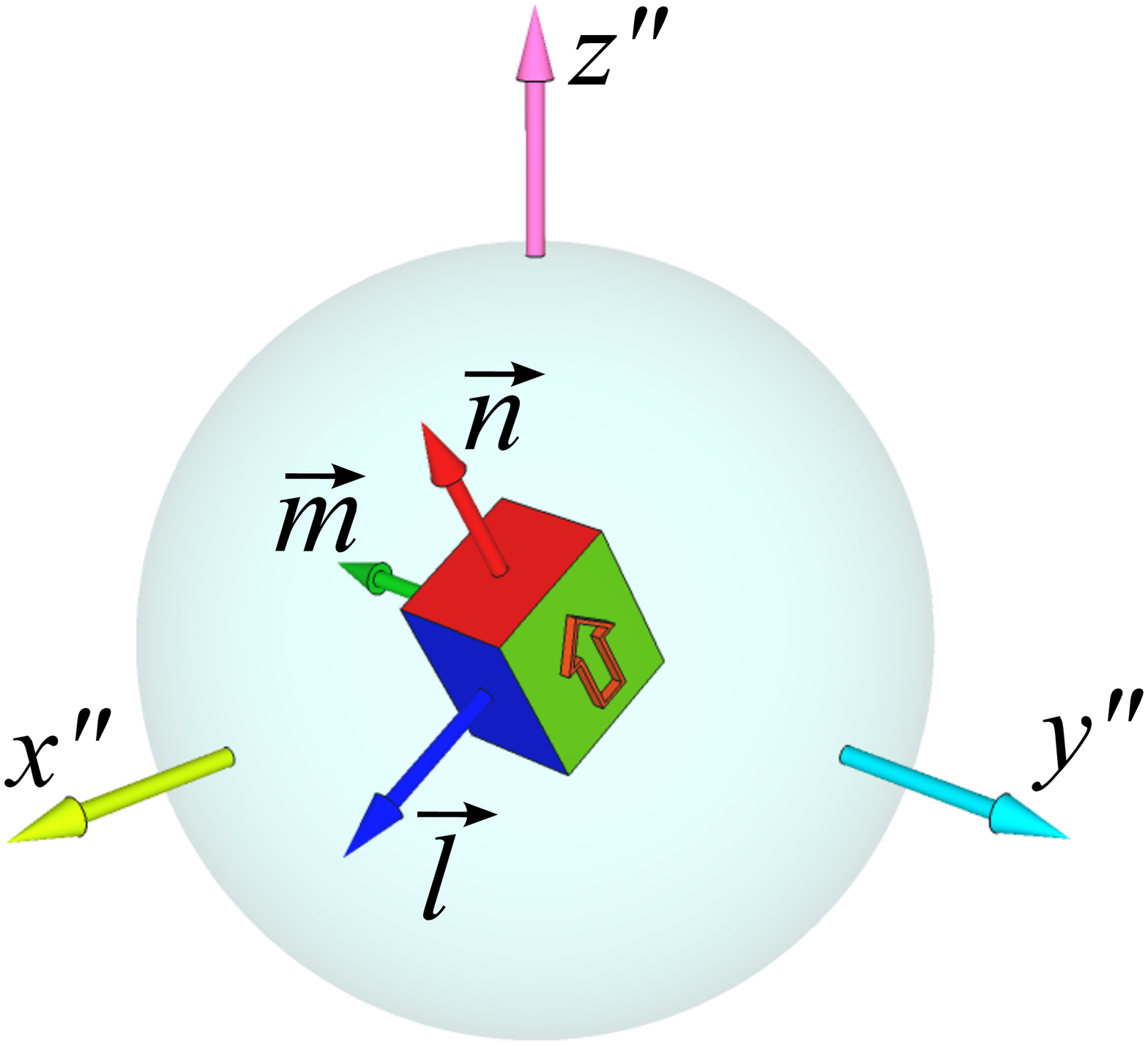}}
      \caption{\small{Double transition to the new system.
      On the left the sequence of rotations of external space 
      relative to the Pauli-operator
      drawn like a colored cube.
      First rotation around the vector $\vec{l}$ 
      is left rotation and defined by the operator $\widehat{U}^+_{\vec l}$
      that gives the system $(x',\,y',\,z')$.
      Further the operator $\widehat{U}_{\vec n}$
      defines right hand rotation around $\vec{n}$ %according to the right hand sqrew rule
      and moves the system to the position $(x'',\,y'',\,z'')$.
      On the right we change the point of view
      to return the laboratory system in initial state
      together with Pauli-operator.
     }}\label{double rotations versus Pauli operator}
   \end{figure}

 Spinors are called as eigenvectors of Pauli-operator 
 which implies their subordinate position: 
 the operator $\sigma_r$ acts to the spin vector $\vec{s}$
 and represents its state as a spinor $\chi_r$. 
 But the formulas 
 (\ref{new definition of Pauli matrices}\,a,\,b,\,c)
 give inverse interpretation ---
 the spin produces own Pauli-vector and is linked with it as a whole.
 We can say even that the Pauli-vector is a kind of form of the same spin.
 It does not matter how the spin is oriented in the laboratory coordinate system 
 but anyway it have <<own opinion>> about the rules of changing in the external space.
 It is easy to understand using the sequence of rotations
 (Fig.~\ref{double rotations versus Pauli operator}).
 In the begining the basic $(\vec{i},\,\vec{j},\,\vec{k})$ of system $(x,\,y,\,z)$
 and basic $(\vec{m},\,\vec{l},\,\vec{n})$ of Pauli-operator are coincide.
 First rotation is performed around the $y$-axis by the operator 
 $\widehat{U}^+_y\equiv\widehat{U}^+_{\vec l}$
 that leads to the system $(x',\,y',\,z')$.
 Further the operator $\widehat{U}_{\vec n}$
 transforms the system to the next position $(x'',\,y'',\,z'')$.
 However the spin and Pauli-vector remained motionless
 and therefore the $\widehat{U}_{\vec n}$ operator
 has the same form like a matrix $u_z$ \eqref{appendix.su2 around z}:
 \begin{equation*}
   \widehat{U}_{\vec n}=
   \begin{pmatrix}e^{-i\tfrac{\varphi}{2}}&0\\0&e^{i\tfrac{\varphi}{2}}\end{pmatrix}=
   \exp\Bigl[-i\dfrac\varphi2\sigma_z\Bigr]\;.
 \end{equation*}
 This rotation is performed not around laboratory axis $z$
 but around the vector $\vec{n}$, \linebreak
 i.e.~own $z$-axis of fermion
 where the spin produces the projection $\sigma_z$ 
 (\ref{new definition of Pauli matrices}\,a).

\pagebreak
   
\begin{figure}[!ht]
     \centering
      \scalebox{.2}{\includegraphics{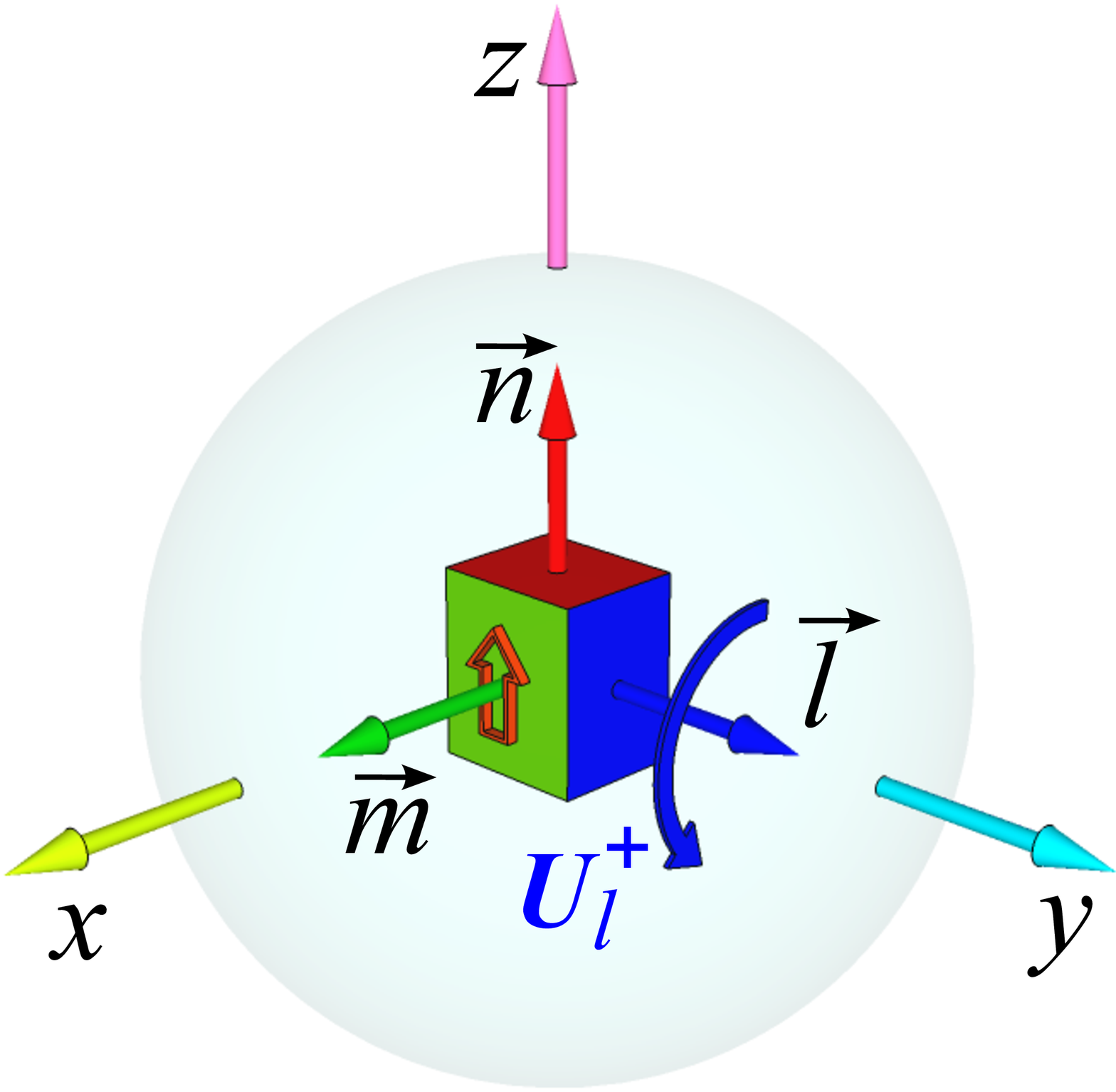}}\;
      \scalebox{.2}{\includegraphics{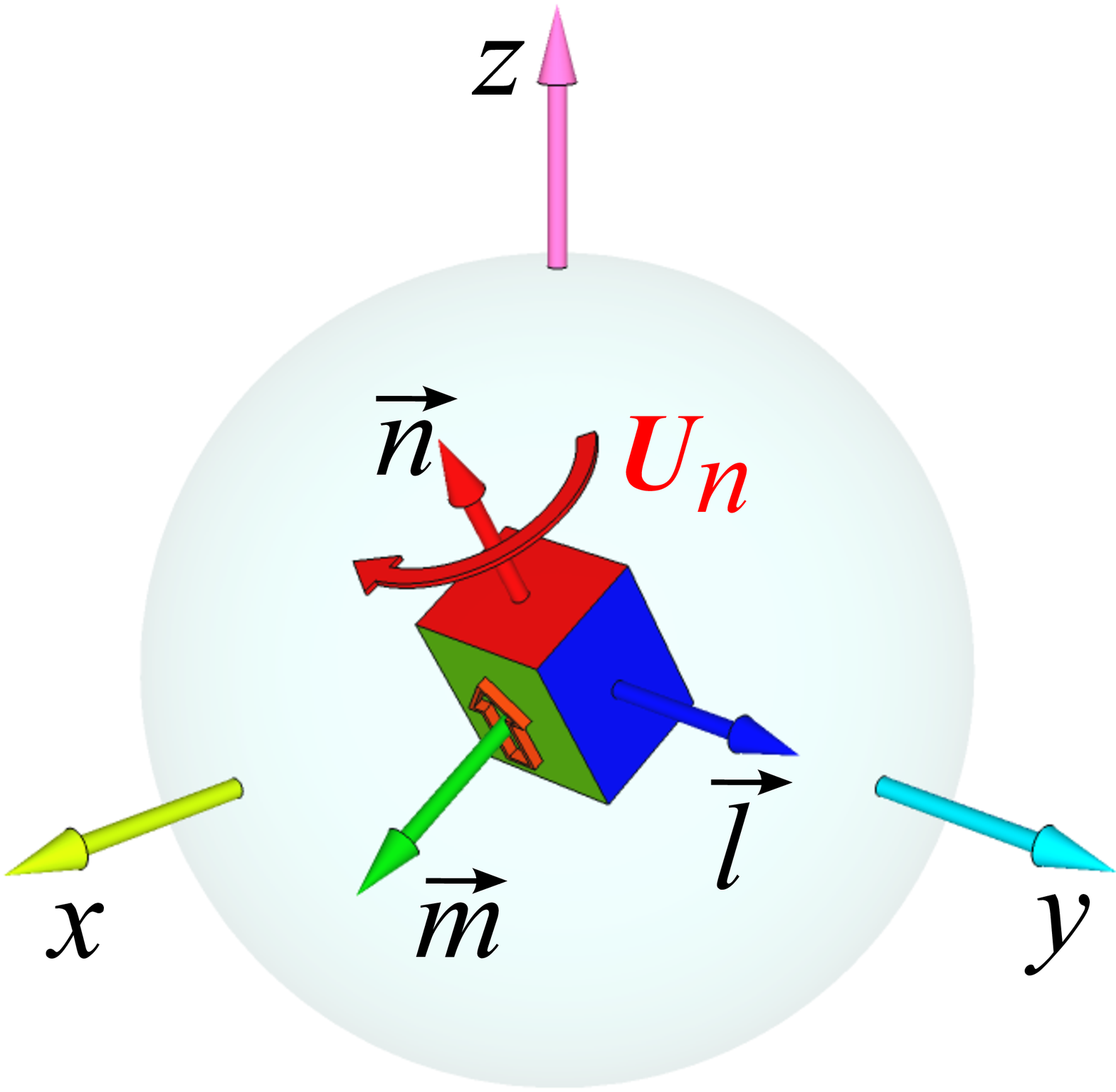}}\;
      \scalebox{.2}{\includegraphics{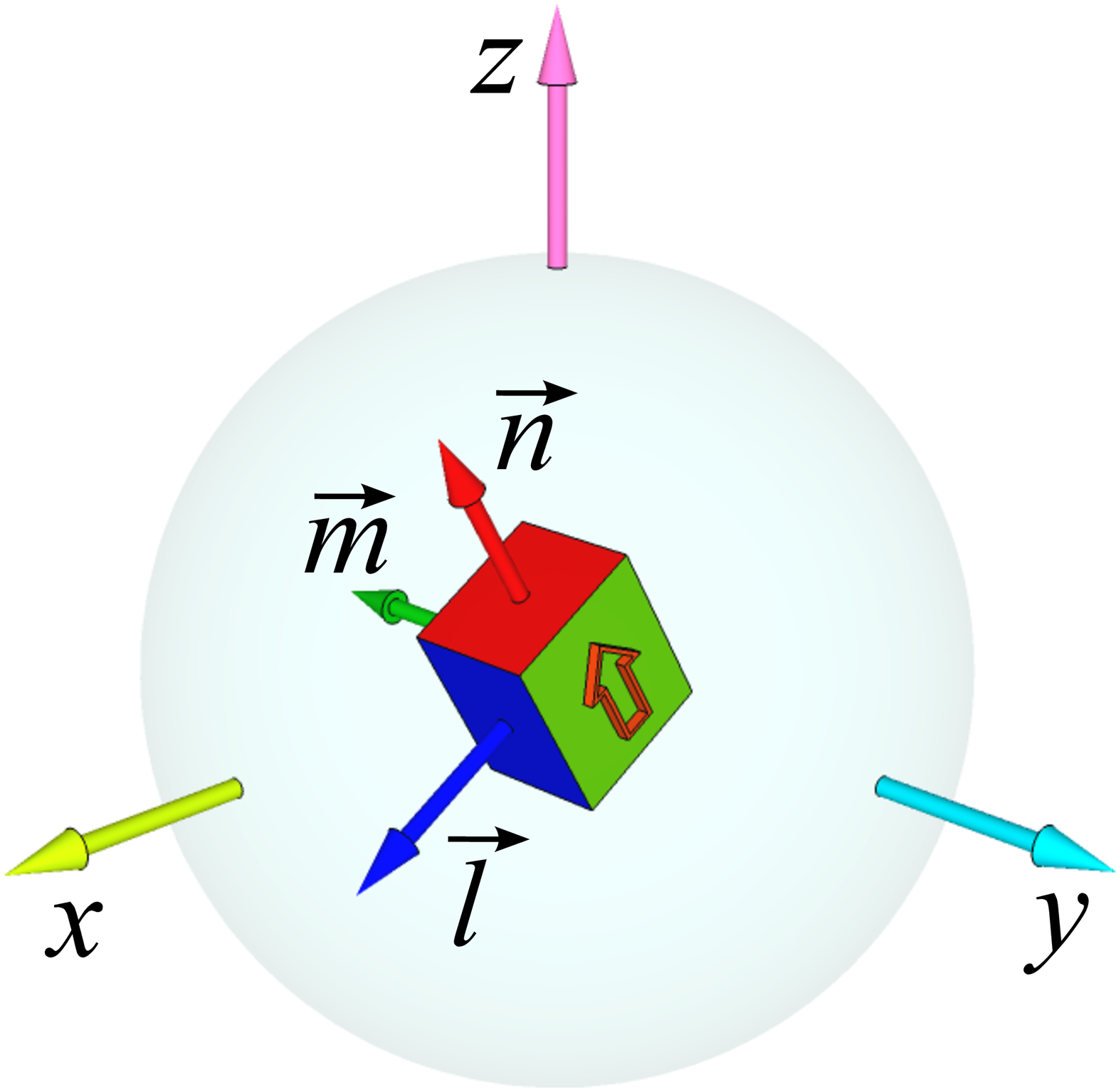}}
      \caption{\small{The transformation of Pauli-operator in the laboratory system $(x,\,y,\,z)$.
      First rotation around the vector $\vec{l}$ to the right 
      is given by the operator $\widehat{U}^+_{\vec l}$.
      On the second picture the left rotation is
      defined by the operator $\widehat{U}_{\vec n}$
      Third picture shows the result of these two operations.
     }}\label{double rotations versus external space}
   \end{figure}

 This sequence of rotations
 (Fig.~\ref{double rotations versus Pauli operator})
 can be presented by the other way
 following to the coordinate system $(x,\,y,\,z)$,
 i.e.~considering it as fixed.
 In this case the state of Pauli-operator is changed
 (Fig.~\ref{double rotations versus external space})
 but its transformations is defined using the same operators
 $\widehat{U}^+_{\vec l}$ and $\widehat{U}_{\vec n}$.
 It is obviously that both pictures \ref{double rotations versus Pauli operator}.2
 and \ref{double rotations versus external space}.3 are identical,
 i.e.~the final state in two different approaches is the same.
 In the books of quantum mechanic we read usualy
 about coordinate system rotation  
 but on the pictures the rotations of operator $\widehat\sigma$ are shown.

\pagebreak
\section{The Bloch Sphere}\label{Appendix.Bloch sphere}
\begin{figure}[!ht]
     \centering
      \scalebox{.2}{\includegraphics{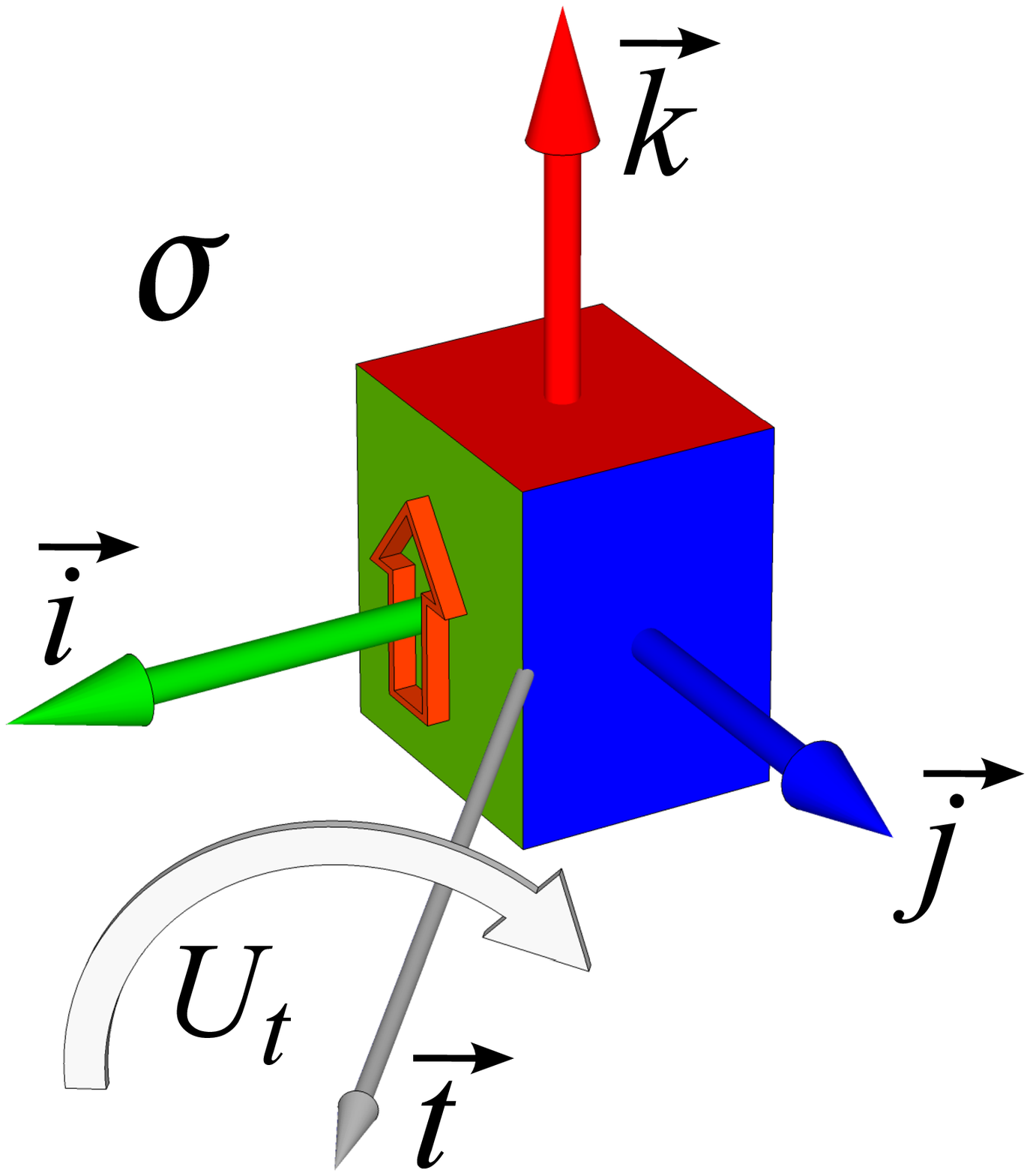}}\qquad
      \scalebox{.2}{\includegraphics{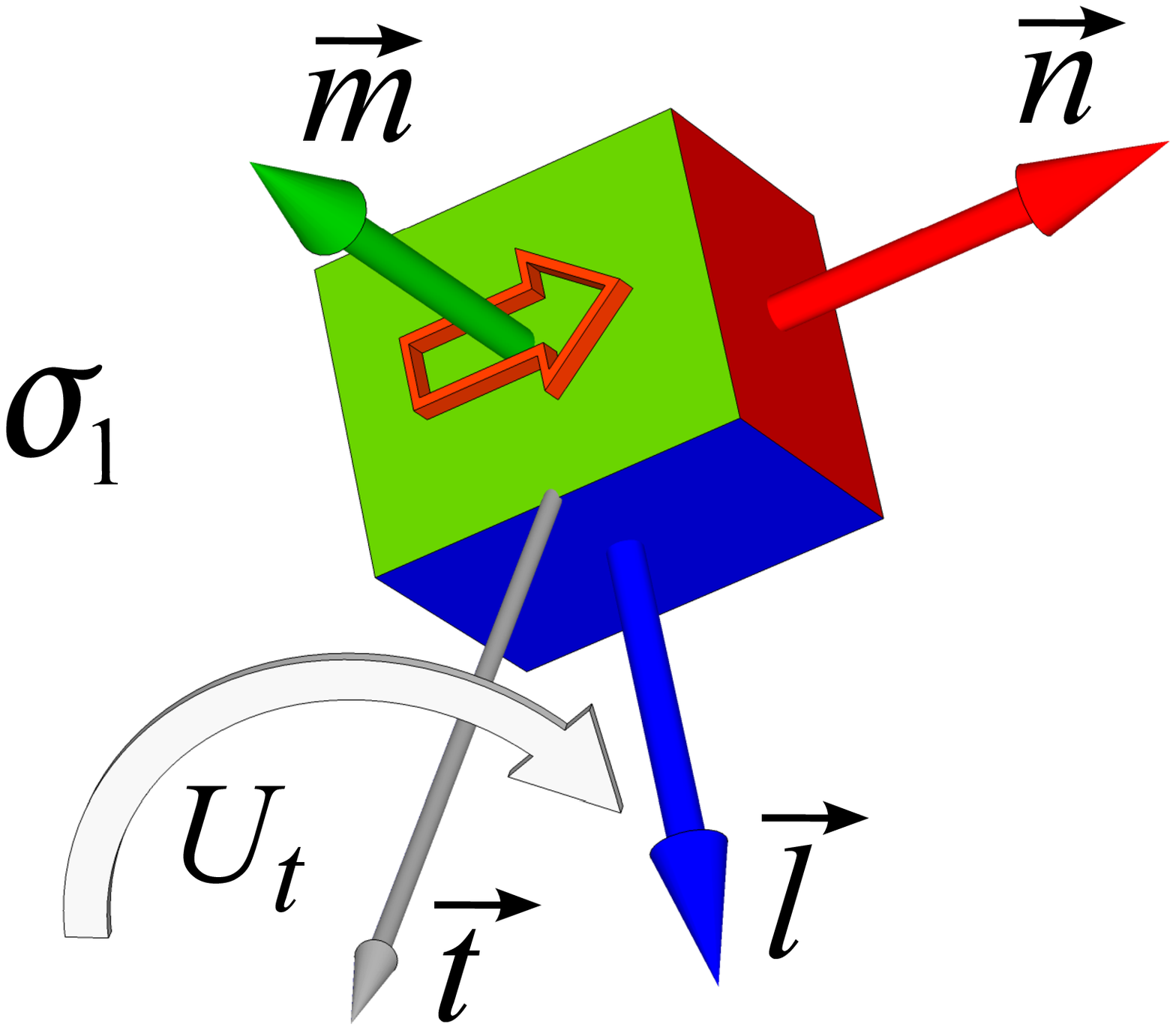}}
      \caption{\small{Symbolic presentation of two Pauli-operators
      $\widehat\sigma$ and $\widehat\sigma_1$
      in the basics $(\vec{i},\,\vec{j},\,\vec{k})$ and
      $(\vec{m},\,\vec{l},\,\vec{n})$ respectively.
      They are related among themselves by the rotation
      around the $\vec{t}$-axis by the angle $\theta$
      that equal to the unitary transformation:
      $\widehat\sigma_1=\widehat{U}_{\vec t}\,\widehat\sigma\widehat{U}^+_{\vec t}$
      where
      $\widehat{U}_{\vec t}=E\cos\frac\theta2-i\sigma_t\sin\frac\theta2$
      and
      $\sigma_t=(\widehat\sigma\cdot\vec{t}\,)=(\widehat\sigma_1\cdot\vec{t}\,)$
      since the axis $\vec{t}$ is the same.
     }}\label{mutual_axize_of_two_operators}
   \end{figure}

  Let us to introduce the concept of global operator $\widehat\sigma$ 
  which is fixed in the laboratory system $(x,\,y,\,z)$.
  Any other spin operator $\widehat\sigma_1$ according to the Euler theorem
  can be obtained from one using a rotation 
  around some axis $\vec{t}$ (Fig.~\ref{mutual_axize_of_two_operators}).

  Let the operators $\widehat\sigma$ and $\widehat\sigma_1$ 
  are defined by the standard way \eqref{theory.Spin Pauli matrices}:
   \begin{equation*}
     \begin{array}{cc}
       \sigma_{x}=\sigma_{1m}=\begin{pmatrix}0&1  \\ 1&0\end{pmatrix}\;,\quad
       \sigma_{y}=\sigma_{1l}=\begin{pmatrix}0&-i \\ i&0\end{pmatrix}\;, \quad
       \sigma_{z}=\sigma_{1n}=\begin{pmatrix}1&0  \\ 0&-1\end{pmatrix}\;.
     \end{array}
   \end{equation*}
   The coordinates of $\widehat\sigma_1$ 
   in the system $(x,\,y,\,z)$ will respect to the formulas
   \eqref{appendix.transition of Pauli-operator}:
   \begin{eqnarray*}
     & (\sigma_{1x},\,\sigma_{1y},\,\sigma_{1z})=
       \widehat{U}_{\vec t}\,(\sigma_{x},\,\sigma_{y},\,\sigma_{z})\widehat{U}^+_{\vec t}\;, \\ [2mm]
     & \widehat{U}_{\vec t}=E\cos\dfrac\theta2-i(\widehat\sigma\cdot\vec{t}\,)\sin\dfrac\theta2\;,
     \quad\textrm{$\widehat{U}_{\vec t}$\, --- left sqrew}.
   \end{eqnarray*}

   For clarity let the projection of spin along the vector $\vec{n}$ equals to $+\frac12\hbar$.
   Its state $\chi_{1n}=\binom10$
   is reflected to the system $(x,\,y,\,z)$
   using inverse trasformaion: $\chi_{1z}=\widehat{U}^+_{\vec t}\binom10=\binom\alpha\beta$
   where $\alpha,\,\beta\in\mathbb{C}$, $|\alpha|^2+|\beta|^2=1$.
   Consider the operator of right rotation around an arbitrary axis $\vec{q}$:
   \begin{eqnarray}
     & \widehat{U}^+_{\vec q}=E\cos\dfrac\theta2+i(\widehat\sigma_1\cdot\vec{q}\,)\sin\dfrac\theta2\;, \\ [2mm]
     & (\widehat\sigma_1\cdot\vec{q}\,) = \sigma_{1m}q_m + \sigma_{1l}q_l + \sigma_{1n}q_n =
       \sigma_{1x}q_x + \sigma_{1y}q_y + \sigma_{1z}q_z = 
       \widehat{U}_{\vec t}\,(\widehat\sigma\cdot\vec{q}\,)\widehat{U}^+_{\vec t}\,.\qquad
   \end{eqnarray}
   For scalar multiplying of spinors $\chi_{1n}$ and $\widehat{U}^+_{\vec q}\chi_{1n}$
   it gives the following:
   \begin{eqnarray}
     & (\chi_{1n})^+\widehat{U}^+_{\vec q}\chi_{1n} = (\chi_{1n})^+\widehat{U}_{\vec t}
       \left(E\cos\frac\theta2+i(\widehat\sigma\cdot\vec{q}\,)\sin\frac\theta2\right)\widehat{U}^+_{\vec t}\chi_{1n} =
       \nonumber \\ [2mm]
     &  =\left(\widehat{U}_{\vec t}^+\chi_{1n}\right)^+\widehat{R}^+_{\vec q}\left(\widehat{U}^+_{\vec t}\chi_{1n}\right)\;, \quad
       \widehat{R}^+_{\vec q}=E\cos\dfrac\theta2+i(\widehat\sigma\cdot\vec{q}\,)\sin\dfrac\theta2\;.
   \end{eqnarray}
   It means that the operator $\widehat{U}^+_{\vec q}$
   in the coordinate system of Pauli-operator $\widehat\sigma_1$
   changes the spinor $\chi_{1n}=\binom10$
   in the same way as the operator $\widehat{R}^+_{\vec q}$
   changes its reflection $\chi_{1z}=\binom\alpha\beta$
   in the system $(x,\,y,\,z)$ of global operator $\widehat\sigma$.
   Thus we can define the trivial rule:
   
  \begin{lemma}[Arbitrary spin-space]\label{appendix.arbitrary spin-space.lemma}
  \setlinespacing{1.66}
    The rotation of spin vector around any axis $\vec{q}$
    carries out identical at any choosing of quantization axis $z$.
  \end{lemma}

  Because the spin transformation depends only
  from the position of rotation axis 
  and from the value of rotation angle
  it allows to use common for all fermions Pauli-operator $\widehat\sigma$,
  \begin{figure}[!ht]
   \quad\centering
   \scalebox{.3}{\includegraphics{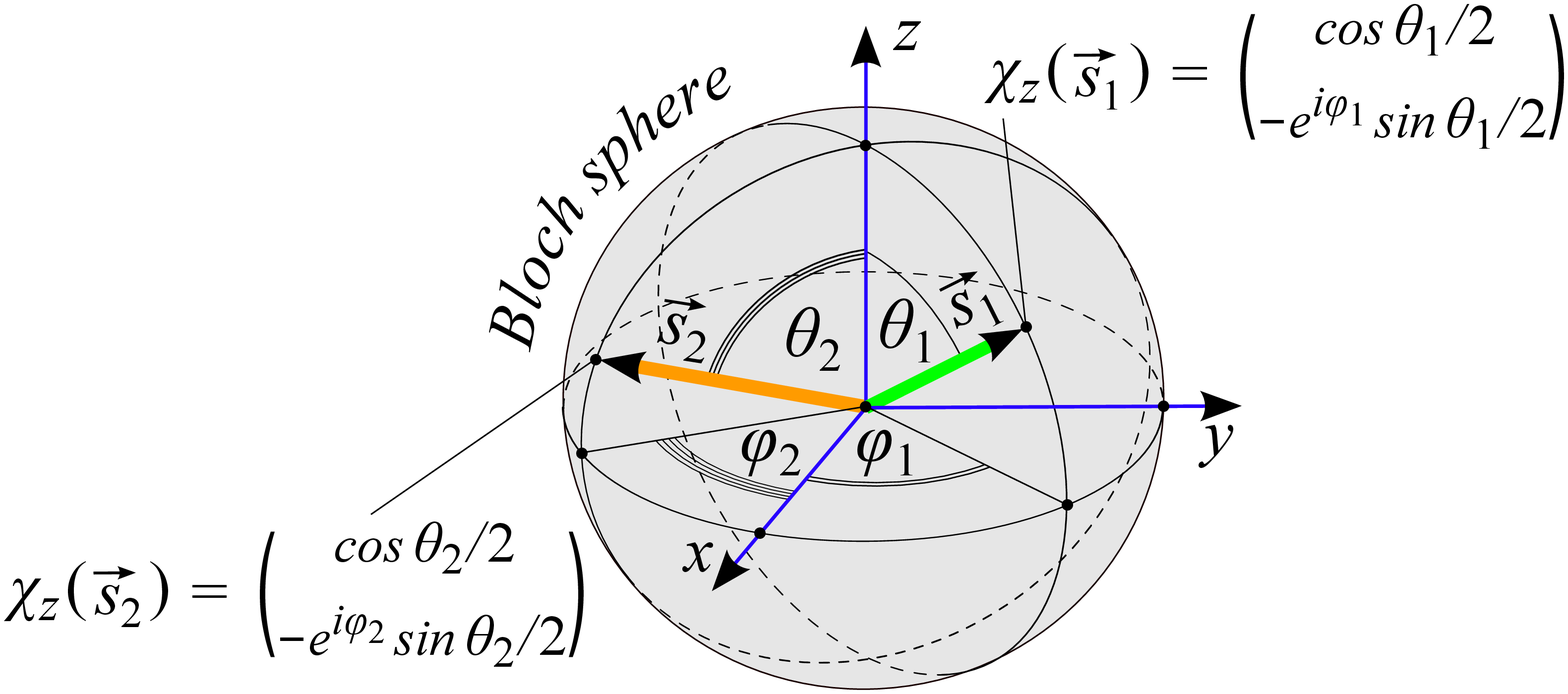}}\qquad
   \caption{\small{The Bloch sphere.
    The spin direction $\vec{s}\equiv\vec{s}(\theta,\varphi)$
    is defined by the angles $\theta$ and $\varphi$.
    For each $\vec{s}$ the point on the Bloch sphere
    corresponds to the single spin state along the axis $z$:
    $\chi_z(\vec{s})=\cos\frac\theta2\binom10-e^{i\varphi}\sin\frac\theta2\binom01$.
    The sign minus before low spinor element 
    arises due to the inverse count of angle $\theta$
    in comparison with the case on Fig.~\ref{theory.Pauli-operator sphere}.
   }}\label{theory.Bloch sphere}
  \end{figure}
  i.e.~we can work in the representation of Bloch Sphere (Fig.~\ref{theory.Bloch sphere}).
  Global space of this {\it referee}\,-operator is a spin space
  in the sense that each spin is transformed here like in its own space,
  i.e.~the right rotation around any axis $\vec{q}$
  is given by the same form of unitary operator
  $\widehat{U}^+_{\vec q}(\varphi)=\exp{(+i\frac\varphi2(\widehat\sigma\cdot\vec q\,))}$.

\pagebreak
\section{Conclusion}
 \begin{enumerate}
  \item
   Wrong interpretation of spin $\frac12$ transformation
   is related with the casus that unitary operator
   $\widehat{U}_{\vec n}(\varphi)=\exp{(-i\frac\varphi2(\widehat\sigma\cdot\vec n\,))}$
   gives the right rotation not for spin vector but for the Stern-Gerlach device.
   Its axis of quantization is directed along the vector $\vec{r}$
   and can moved on the $\Sigma$-sphere (Fig.~\ref{theory.Pauli-operator sphere})
   at the fixed spin projection $s_z=+\frac12\hbar$.
%   In the SU(2) algebra 
   The Stern-Gerlach device can be associated
   with the operator $(\widehat\sigma\cdot\vec{r}\,)$
   and its spinor %eigenvector 
   $\chi_r(s_z=+\frac12\hbar)=\cos\frac\theta2\binom10+e^{i\varphi}\sin\frac\theta2\binom01$
   defines the amplitudes $\cos\frac\theta2$ and $e^{i\varphi}\sin\frac\theta2$
   of projections $s_r=+\frac12\hbar$ и $-\frac12\hbar$. %по направлению $\vec{r}$,
  \item
   Each fermion in own coordinate system $(x_i,\,y_i,\,z_i)$
   produces own Pauli-vector $\widehat\sigma_i$.
   The inverse unitary operator
   $\widehat{U}^+_{\vec n}(\varphi)=\exp{(+i\frac\varphi2(\widehat\sigma_i\cdot\vec n\,))}$
   performs the right rotation of spin vector in the external space
   from which this fermion seems to be repelled
   (Fig.~\ref{double rotations versus external space}).
  \item
   If in the laboratory system $(x,\,y,\,z)$ we define
   the {\it referee}\,-operator $\widehat\sigma$
   which respects to the fixed position of Stern-Gerlach device
   the direction of spin is considered as free
   and defined by the vector $\vec{r}$ on the Bloch Sphere
   (Fig.~\ref{theory.Bloch sphere}).
   In this case the spin is quantized along the axis $z$ 
   and its state is presented by the spinor 
   $\chi_z(s_r=+\frac12\hbar)=\cos\frac\theta2\binom10-e^{i\varphi}\sin\frac\theta2\binom01$
   where the $\cos\frac\theta2$ and $-e^{i\varphi}\sin\frac\theta2$
   are the amplitudes of projection $s_z=+\frac12\hbar$ and $-\frac12\hbar$.
   The rotation of spin vector is given by the unitary operator
   $\widehat{U}^+_{\vec n}(\varphi)=\exp{(+i\frac\varphi2(\widehat\sigma\cdot\vec n\,))}$.   
 \end{enumerate}

%\fi

%\bibliographystyle{amsplain}
%\bibliographystyle{my-ieeetr}
\bibliographystyle{ieeetr}
\bibliography{./bib/shindin_dubna_bib/MyThesisBib}

% ----------------------------------------------------------------

\end{document}